\documentclass[prb,twocolumn,showpacs,preprintnumbers,superscriptaddress]{revtex4}%

\usepackage{graphicx}
\usepackage{dcolumn}
\usepackage{bm,hyperref}
\usepackage{amsmath}
\usepackage{amsfonts}
\usepackage{amssymb}%
\providecommand{\U}[1]{\protect\rule{.1in}{.1in}}
\begin{document}

\title{High symmetry versus optical isotropy of a negative index metamaterial}
\author{Christoph Menzel}
\affiliation{Institute of Condensed Matter Theory and Solid State
Optics, Friedrich-Schiller-Universit\"at Jena, Max-Wien-Platz 1,
D-07743 Jena, Germany}

\author{Andrei Andryieuski}
\affiliation{DTU Fotonik - Department of Photonics Engineering,
Technical University of Denmark, {\O}rsteds pl. 343, DK-2800 Kgs.
Lyngby, Denmark}

\author{Carsten Rockstuhl}
\affiliation{Institute of Condensed Matter Theory and Solid State
Optics, Friedrich-Schiller-Universit\"at Jena, Max-Wien-Platz 1,
D-07743 Jena, Germany}

\author{Rumen Iliew}
\affiliation{Institute of Condensed Matter Theory and Solid State
Optics, Friedrich-Schiller-Universit\"at Jena, Max-Wien-Platz 1,
D-07743 Jena, Germany}

\author{Radu Malureanu}
\affiliation{DTU Fotonik - Department of Photonics Engineering,
Technical University of Denmark, {\O}rsteds pl. 343, DK-2800 Kgs.
Lyngby, Denmark}

\author{Falk Lederer}
\affiliation{Institute of Condensed Matter Theory and Solid State
Optics, Friedrich-Schiller-Universit\"at Jena, Max-Wien-Platz 1,
D-07743 Jena, Germany}

\author{Andrei V. Lavrinenko}
\affiliation{DTU Fotonik - Department of Photonics Engineering,
Technical University of Denmark, {\O}rsteds pl. 343, DK-2800 Kgs.
Lyngby, Denmark}

\begin{abstract}
Optically isotropic metamaterials (MMs) are required for the
implementation of subwavelength imaging systems. At first glance
one would expect that their design should be based on unit cells
exhibiting a cubic symmetry being the highest crystal symmetry. It
is anticipated that this is a sufficient condition since it is
usually assumed that light does not resolve the spatial details of
MM but experiences the properties of an effective medium, which is
then optically isotropic. In this work we challenge this
assumption by analyzing the isofrequency surfaces of the
dispersion relation of the split-cube in carcass (SCiC) negative
index MM. We show that this MM is basically optically isotropic,
but not in the spectral domain where it exhibits negative
refraction. The primary goal of this contribution is to introduce
a tool that allows to probe a MM against optical isotropy.
\end{abstract}

 \maketitle

\section{Introduction}
Driven by the desire and the opportunity to have optical materials
with tailored properties at hand, the field of metamaterials (MMs)
attracts a steady increasing share of research interest. To
loosely define the field, one may understand MMs as artificial
structures made of subwavelength unit cells , called metaatoms.
They predominantly affect the light propagation by a careful
choice of their geometry and their arrangement; and not by the
intrinsic properties of the materials they are made of. MMs allow
to tailor the flow of light well beyond what would be possible
with naturally occurring materials by mimicking unprecedented
optical properties. A large variety of new optical phenomena were
predicted and experimentally proven on the base of MMs, where
negative refraction attracted potentially the largest amount of
interest\cite{Shelby2001,Smith2004,Dolling2006Science,Soukoulis2007Science,Lezec2007,Yao2008}.
Metamaterials providing a negative effective
refractive index are the essential ingredient to fabricate a
perfect lens with an optical resolution limit well below Abbe's
prediction\cite{Pendry2000}. However, and this is the major requirement
currently not met by most of present MMs, this effective
refractive index can only be meaningfully introduced if the
dispersion relation is isotropic, i.e. for lossless media the isofrequency surface
of the dispersion relation in 3D k-space must be spherical. Moreover, this
isotropy has to hold also for evanescent waves, since they carry
the subwavelength information for the desired super resolution. To overcome this
obstacle various approaches for obtaining isotropic MMs have been put forward
\cite{Balmaz2002,Kussow2008,Vendik2006,Jelinek2006,Baena2006,Simovski2003,Verney2004}.

A systematic approach to design an isotropic magnetically active
MM was proposed by Baena et al. \onlinecite{Baena2007}. The first step is to
choose a highly symmetric MM. Such MM is constructed based on a
cubic unit cell arranged at a cubic lattice. It is evident that
MMs with this high symmetry behave isotropically in the
quasi-static limit. Frequently the optical properties of
such MMs have only been investigated for normal incidence (zero
transverse wave vector) and then extrapolated towards finite
transverse wave vectors, assuming isotropic medium response due to
the symmetry of the structure.

However, and this is crucial, even if the structure is operated in
the subwavelength domain, the required optical isotropy is not
straightforward, since the typical resonance wavelengths, evoking
magnetic effects,  are comparable to the structure size. There are
only a few attempts where the MMs are probed by obliquely incident
fields and for different polarizations, mostly restricted to
transmission and reflection measurements
\cite{Koschny2005,Enkrich2005,Gundogdu2006,Menzel2009APL,Menzel2009JOSAB}.
Since even for an
isotropic medium the reflection and transmission coefficients
evidently depend on the angle of incidence, it is hardly possible
to draw conclusion with respect to the isotropic behavior of the
respective MM. To ultimately verify whether a highly symmetric
metamaterial behaves optically isotropic, it is necessary to
calculate the dispersion relation $\omega  = \omega \left(
{{k_{\rm{x}}},{k_{\rm{y}}},{k_{\rm{z}}}} \right)$ where its
iso-frequency surface  ${k_{\rm{z}}} = {k_{\rm{z}}}\left(
{{k_{\rm{x}}},{k_{\rm{y}}},\omega  = {\rm{const}}.} \right)$
governs the diffraction and refraction properties of the MM
\cite{Paul2009}. Here, $\omega$, ${k_{{\rm{x}}{\rm{,y}}}}$, and
${k_{\rm{z}}}$ are the frequency, the transverse and the
longitudinal wave vector components, respectively, where the
latter is frequently termed propagation constant. The aim of this
work consists just in evaluating this iso-frequency surface.

For computing iso-frequency surfaces of a three-dimensional MM we
take the complex permittivity of the material into account. This
is crucial because most the band structure solvers assume lossless
media which is certainly incorrect for metamaterials. We use
instead a plane wave expansion technique that solves Maxwell's
equations in the frequency domain for the periodic structure. The
technique solves the respective eigenvalue problem for the
generally complex propagation constant of the Bloch modes\cite{Li2003PRE}.
Computing the propagation constant as a function of the transverse
wave vector for a fixed frequency provides the iso-frequency
surfaces. By using this method we specifically show for the
split-cube in carcass (SCiC)-MM with cubic symmetry that its
optical response is isotropic only at low frequencies. By contrast
this isotropy disappears in the spectral domain where the
propagation constant, and hence the effective refractive index is
negative.

After having introduced the system under consideration in Sec. II,
we show in Sec. III that the proposed MM fulfills all requirements
usually imposed for its homogenization. In Sec. IV we discuss the
iso-frequency surface for the negative refraction regime and
introduce a simple measure to quantify the optical isotropy across
the entire sub-wavelength frequency domain.

\section{Split-cube in carcass design}
For our investigations we use the split-cube in carcass
structure, which is a simplified version of the split-cube in cage
structure \cite{Andryieuski2009JOA} containing less number of fine
details. Using the nested structures approach the SCiC is designed such,
that one element of the unit cell that exhibits magnetic properties
is inserted into another element which shows dielectric resonances
\cite{Andryieuski2009JEOS}. The SCiC unit cell consists of two silver parts
embedded in silica ($\varepsilon=2.25$). Silver is regarded as Drude metal
with a plasma frequency of $\omega_\mathrm{p}=1.37\cdot10^{16}
\mathrm{ rad/s}$ and collision frequency
$\omega_\mathrm{c}=8.5\cdot10^{13}1/\mathrm{s}$ \cite{Dolling2006OL}.
These parameters were adjusted to emulate the real experimental
situation in simulations.
\begin{figure}[h]
\centering
\includegraphics[width=84mm,angle=0] {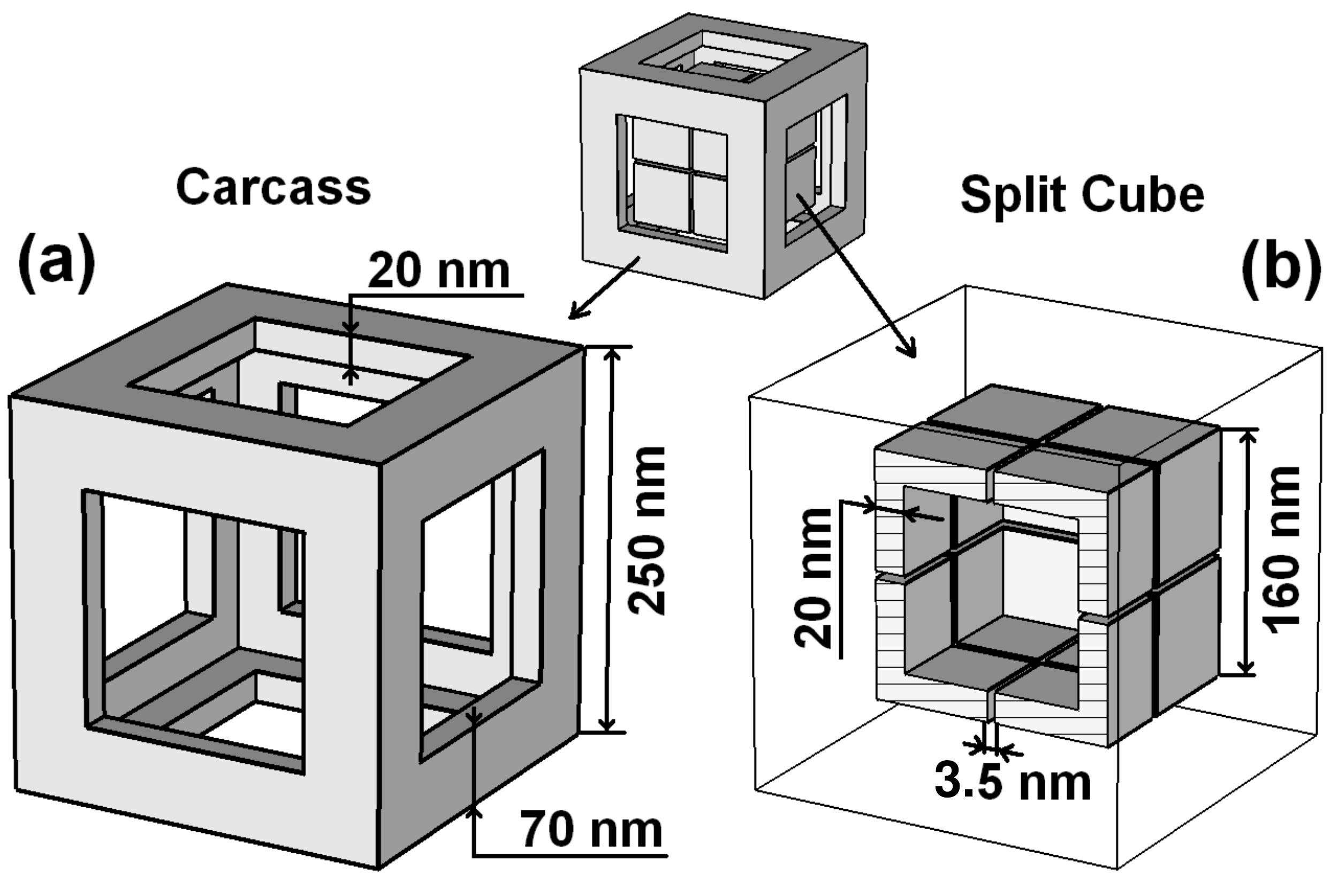}
\caption[Submanifold] {(color online) Schematic of the Split Cube
in Carcass unit cell. The Carcass (a) serves to provide a
plasma-like optical response resulting in a negative effective
permittivity. The Split Cube cut (cut is shown for clarity) (b)
provides an artificial magnetic response.} \label{FIG_KOS}
\end{figure}

The outer part, the Carcass, is a three-dimensional wire medium
\cite{Pendry1996,Belov2003} and provides a dielectric resonance ('negative permittivity'
$\varepsilon$). The inner part, the Split Cube, is a hollow cube
with slits in the middle of the facets. It is the logical 3D
extension of the symmetric split ring resonator \cite{Hardy1981,Linden2006Science,Penciu2008}
concept and provides a magnetic resonance ('negative permeability'
$\mu$). The details regarding the structure's sizes are indicated
in Fig.~\ref{FIG_KOS}.

The SCiC was chosen because it has the highest possible symmetry
for a periodically arranged MM. In particular, it is
mirror-symmetric with respect to three orthogonal axes. This
excludes any effects resulting from first order spatial dispersion
like chirality \cite{Koschny2005,Baena2007,SerdyukovBook}. Due to the cubic
symmetry the three main propagation directions are equivalent.
Hence, the optical response of the SCiC is supposed to be
described by scalar effective material parameters in the
quasistatic limit, i.e. the optical response should be isotropic.
From these symmetry considerations one usually draws the
conclusion that the SCiC might be an ideal candidate for an
isotropic negative index metamaterial.

\section{Homogeneity}
To describe a MM as an effectively homogeneous medium, it is
necessary to assure at first that light propagation within the MM
is exclusively governed by a single Bloch mode. To be sure
that this requirement is met, we also have to prove that the
incident light field couples only to this particular mode. This is
done by comparing the effective plane wave propagation constant,
as retrieved from reflection and transmission data of a finite
structure, with that obtained from the dispersion relation of the
lowest order Bloch mode\cite{Rockstuhl2008PRB}.

We will determine the effective propagation constant in the SCiC
by using the S-parameter retrieval method \cite{Smith2002,Menzel2008PR}.
The complex reflection and transmission coefficients are numerically
calculated  by the Fourier Modal Method \cite{LiFMM} where $31\times 31$
Fourier orders were retained to achieve convergent results. By
inversion of these scattering data the effective propagation
constant $k=k_\mathrm{z}$ and the effective impedance $Z$ can be
determined. To certify that such an assignment of effective
parameters to structures composed of only a small number of
functional layers is meaningful and that the homogenization is
valid, it is necessary to investigate the convergence of the
parameters towards their bulk values \cite{Rockstuhl2007PRB},
therefore we determine the propagation constant of the Bloch eigenmodes
of the infinite structure as well.

This is done by calculating the T-matrix for a single period and solving the eigenvalue problem
$$\hat{T}\begin{pmatrix}E\\H\end{pmatrix}=\exp(ik\Lambda)\begin{pmatrix}E\\H\end{pmatrix}$$
to obtain the propagation constant $k=k_{\mathrm{Bloch}}$ of the
Bloch modes, where $\Lambda$ is the period in the main propagation
direction.

\begin{figure}[h]
\centering
\includegraphics[width=84mm,angle=0] {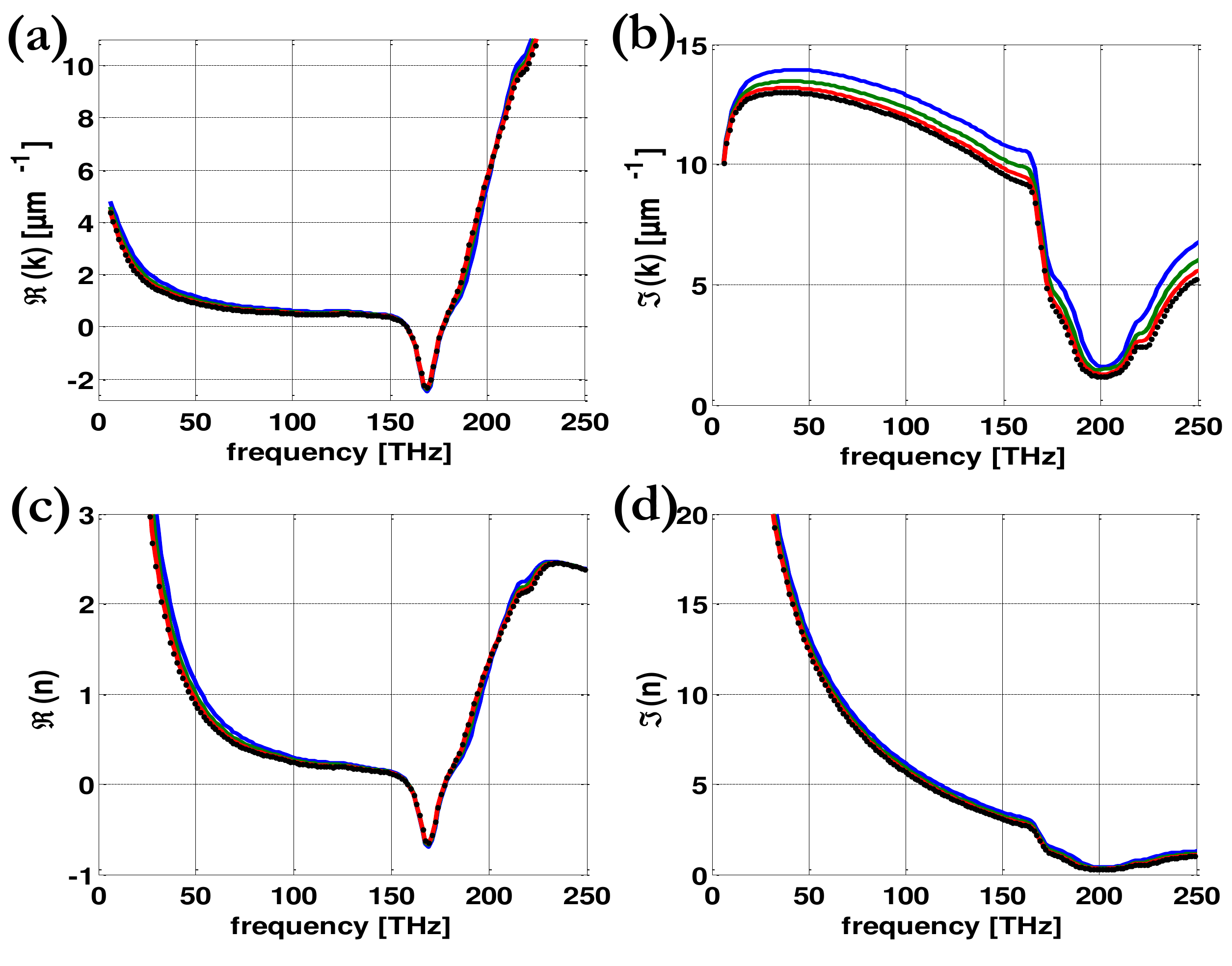}
\caption[Submanifold] {(color online) Real (a) and imaginary (b)
parts of the propagation constants and effective refractive
indices (c) and (d) for the effectively homogeneous finite and the
infinite periodic structures at normal incidence. The results for
the finite structure (solid lines) are given for an increasing
number of layers (1 layer - blue, 2 layers - green, 5 layers -
red). The dotted black line corresponds to the values obtained
from the Bloch mode with the smallest losses.} \label{FIG_KOS0}
\end{figure}

The results for the real and imaginary part of both propagation
constants (finite homogeneous and infinite periodically structured
media) and the formally introduced effective refractive indices
$n=ck/\omega$ are shown in Fig. \ref{FIG_KOS0}.

For the periodic bulk material we get in general the propagation
constants for an infinite number of Bloch modes (which remains
nevertheless finite because of the numerical truncation of the
number of plane waves retained in the plane wave expansion) but
only the zeroth order mode, i.e. the one with the smallest
imaginary part is shown. Clearly the values obtained from the
finite effectively homogeneous structure are converging
astonishingly fast towards the values for the periodic bulk
medium. Thus, two important conclusions can be drawn, firstly the
homogenization of the periodic MM is feasible and secondly the
optical response of a SCiC-MM consisting of only a few layers equals
that of a bulk medium.

The design purpose of the SCiC is a negative effective propagation
constant, and thus an negative effective  index of refraction
which is clearly achieved for frequencies around $170$ THz (Figure
of Merit $-\Re(n)/\Im(n)=0.35$). Note that the effective
permeability is dispersive but positive in the investigated
frequency range. Hence, the SCiC is a single negative MM resulting
in the fairly small Figure of Merit. For frequencies less than
$150$ THz the SCiC is rather a strong absorber due to the large
metal fraction.

Since for the scattering problem at the finite system the coupling
to different Bloch modes was rigorously considered, we can
conclude that the propagation of light through the structure as
well as the coupling process is almost entirely dictated by the
fundamental Bloch mode only, again underlining the validity of the
homogenization procedure. This is clear as the propagation
constant values for the finite and the infinite structures
coincide. Although not shown here for the sake of brevity the
propagation constants for oblique incidence are also rapidly
converging towards the bulk values. Also the effective impedance
is converging for normal incidence as well as for oblique
incidence.

Hence, we can fully rely on describing light propagation in terms
of the dispersion relation assuming that light will couple to this
fundamental mode only. Note that from the results above we can
also conclude that the SCiC fulfills all requirements for the
homogenization of the MM, namely the structure is subwavelength
(compared to the wavelength of the environment) and light
propagation inside the structure is determined by a single Bloch
mode to which an external light field predominantly couples.

\section{Dispersion relation and optical isotropy}
\begin{figure}[h]
\centering
\includegraphics[width=84mm,angle=0] {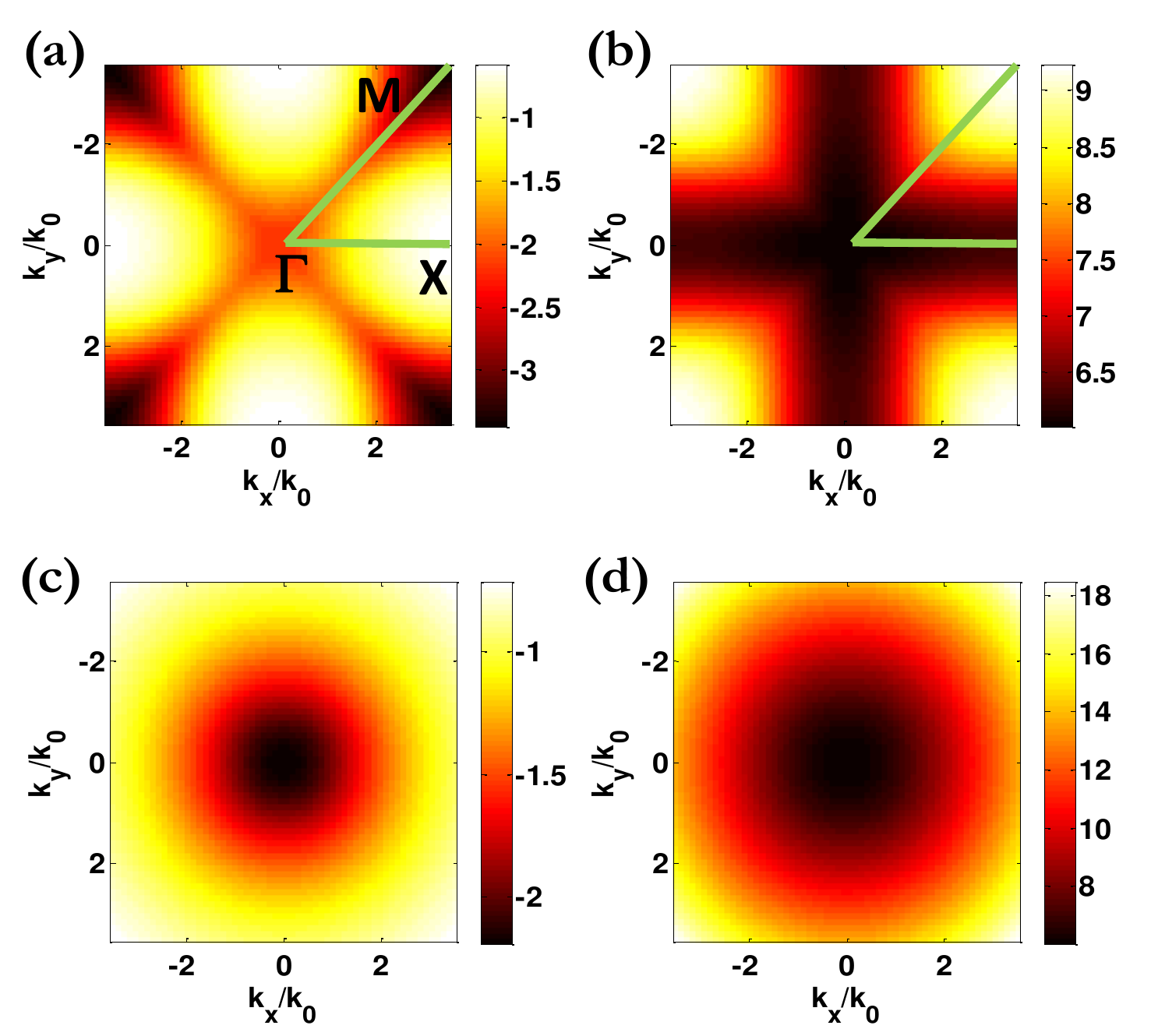}
\caption[Submanifold] {(color online) Real (a) and imaginary (b)
part of the propagation constant of the lowest order Bloch mode in
the first Brillouin zone. Real (c) and imaginary (d) part of the
propagation constant of an isotropic medium with the same
refractive index as the SCiC at normal incidence.}
\label{FIG_KOS1}
\end{figure}

The main property we are interested in is the optical isotropy of the SCiC,
in particular in the spectral region around $170$ THz where the propagation
constant is negative. To judge this we will proceed in calculating the
iso-frequency surface.\\
Due to the symmetry of the structure the complete iso-frequency
surface can be constructed by only calculating the dispersion
relation in the irreducible Brillouin zone. The real and imaginary
part of the propagation constant for a fixed frequency of 170 THz
are shown in Fig.~\ref{FIG_KOS1}(a) and (b). For comparison we
show in Fig.~\ref{FIG_KOS1}(c) and (d) the analytically calculated
iso-frequency surface of an isotropic medium, which has the same
refractive index as SCiC at normal incidence.

At first we notice  that both the real and the imaginary part of
the propagation constant of the MM are not rotationally symmetric
and monotonically changing with increasing absolute value of the
tangential component $k_t$ as for the isotropic medium. In
particular, the dependency on the transverse wave vector is
tremendously different when compared to the isotropic medium.
While the real part is monotonically increasing with $k_t$ in the
$\Gamma\mathrm{X}$-direction it is non-monotonously decreasing in
the $\Gamma\mathrm{M}$-direction. Also the imaginary part is
strongly increasing in the $\Gamma\mathrm{M}$-direction and only
slowly varying in the $\Gamma\mathrm{X}$-direction. Hence, any
formally introduced effective refractive index would explicitly
depend on the transverse wave vector ${{\bf{k}}_ \bot
}=(k_x,k_y)$. Therefore, the introduction of a global effective
refractive index is pointless since no additional information is
obtained. Nevertheless for paraxial wave propagation the
introduction of a local effective refractive index is feasible.
Near the $\Gamma$-point the isofrequency surface is approximately
spherical where the validity of this approximation strongly
depends on the frequency and the wavelength to cell size ratio as
discussed later in detail.

It should be mentioned that the choice of real valued $k_x$ and
$k_y$ is arbitrary to a certain extent. In lossy media like
metamaterials, in general, the complex nature of the dispersion
relation cannot be neglected. Hence the dispersion relation could
also be calculated for complex valued $k_t$. On the other hand it
is pointless to provide these values as they are not accessible in
any experiment. The tangential wave vector components are
continuous at boundaries and only real valued plane wave solutions
can exist in free space, therefore this choice reflects
experimental constraints.

\begin{figure}[h]
\centering
\includegraphics[width=84mm,angle=0] {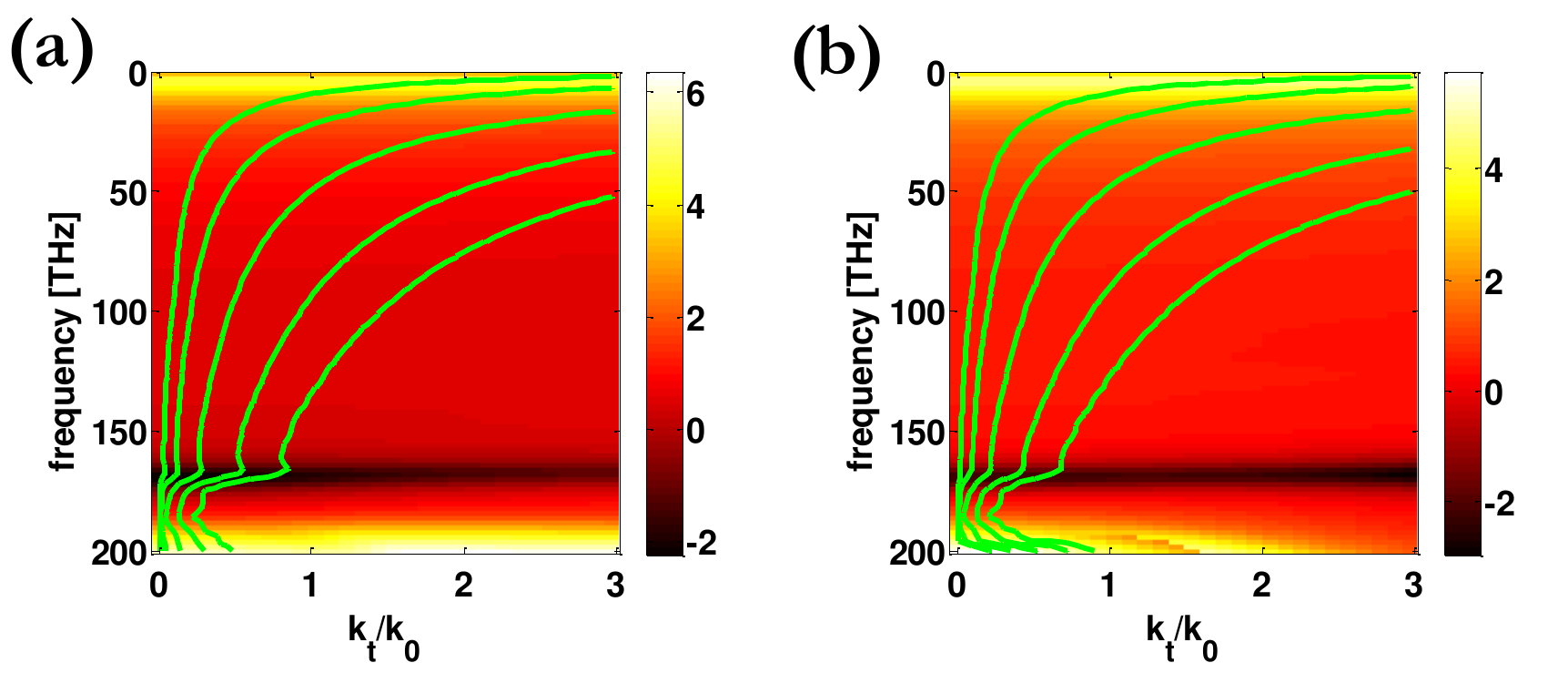}
\caption[Submanifold] {(color online) (a) Real part of the
propagation constant as a function of the frequency and the angle
of incidence where the plane of incidence is parallel to the
coordinate axes. (b) The same as (a) with plane of incidence
rotated by $\pi/4$, i.e. the $\Gamma\mathrm{M}$ direction. The
green iso-error lines correspond to the relative deviations
$\Delta\in [0.02,0.1,0.5,2,5]\%$ (from left to right).}
\label{FIG_KOS2}
\end{figure}

To more quantitatively evaluate the optical isotropy of the MM, we
monitor in the following the relative deviation between the
numerically obtained propagation constant and the ideal spherical
isofrequency surface. To quantify this deviation we assume that
the effective refractive index at normal incidence is valid also
for any other propagation direction. From symmetry considerations
and the exemplary results in Fig.~\ref{FIG_KOS1} we conclude that
it is sufficient to investigate the dependency in the high
symmetry $\Gamma\mathrm{X}$ and $\Gamma\mathrm{M}$ direction
assuming that for a fixed value of $|k_\mathrm{t}|$ these points
are extremes of the iso-frequency surface. In Fig.~\ref{FIG_KOS2}
the real part of the propagation constant is shown as a function
of the frequency and of the tangential wave vector component
$k_\mathrm{t}/k_0$, where $k_0$ is the free space propagation
constant. The edge of the first Brillouin zone for the largest
possible frequency (200THz) would be at
$k_\mathrm{t}/k_0\approx3\,$ because of
\[\frac{k_{\mathrm{t}}}{k_0}=\frac{\pi}{ak_0}=\frac{\lambda(200\textrm{Thz})}{2a}=\frac{2\cdot
10^{6}}{200\cdot 10^{12}}\textrm{c}\approx 3.\] The most important
frequency region here is the black domain of negative refraction
where the Split Cube provides the artificial magnetic response.
Also the iso-error lines for the relative error
\[\Delta  = \left| {\frac{{|{k_{\text{z}}}| - |{k_{\text{z}}}{|_{{\text{ideal}}}}}}
{{|{k_{\text{z}}}{|_{{\text{ideal}}}}}}} \right|\] with
${|k_{\text{z}}|}{_{{\text{ideal}}}}=|\sqrt{k_z^2(k_t=0)-k_t^2}|$
are given for several values as green lines. The quantity $\Delta$
is a measure for the relative deviation of the modulus of the
propagation constant for oblique incidence from that for normal
incidence. We have taken the moduli because all quantities are
complex-valued. The iso-error lines in the non-resonant regime
($\omega\lessapprox 150\,\mathrm{THz}$) scale approximately with
$1/\omega\propto \lambda$. This results from the fact that the
homogeneous medium approximation improves with a decreasing unit
cell size to wavelength ratio $a/\lambda$. For frequencies larger
than $160\,\mathrm{THz}$ the structure becomes resonant and an
abrupt change of the iso-error lines is clearly observable.

Obviously in the resonant regime the MM can be only considered
isotropic if the ratio $a/\lambda$ is much less than in the
non-resonant regime. In general the situation  is  identical for
both planes of incidence ($\Gamma\mathrm{X}$- and
$\Gamma\mathrm{M}$-direction). However, for larger frequencies the
deviation of the calculated iso-frequency lines to the ideal ones
is slightly smaller in $\Gamma\mathrm{M}$- than in
$\Gamma\mathrm{X}$-direction but not significantly.

Note that the introduction of the quantity $\Delta$ is only one
option to quantify the deviation of the pertinent isofrequency
surface from the ideal one. Other measures are also possible, where,
for example, the deviation of the length of the actual wavevector
compared to the length of the wavevector in an isotropic medium
yields qualitatively the same results.

Before we proceed to discuss the results, some remarks concerning
the investigated parameter space are in order. Here, the
dispersion relation is not calculated up to the edge of the
Brillouin zone, but for fixed tangential wave vectors which
translate for propagating waves into certain angles of incidence.
This is more useful since for very low frequencies it is not
possible to provide incident plane waves with tangential wave
vector components $k_t$ that are in the order of $\pi/a$ where $a$
is the lattice constant. Furthermore, plane waves with wave
vectors in the dimension of the lattice vector will always sense
the details of the periodicity. In this case the issue of an
optically isotropic structure is pointless. Of course, also for
small frequencies the iso-surface are not rotationally invariant
but these domains are in general not accessible to the experiment
and are negligible. Assuming the outer medium to have a
permittivity $\varepsilon$ then the line given by
$k_t/k_0=\sqrt{\varepsilon}$ corresponds to grazing incidence and
to the angular domain accessible from free space.

Up to here we have only discussed the properties of the
fundamental Bloch mode with the smallest imaginary part since this
mode essentially dictates light propagation. While considering the
coupling of an external field to this Bloch mode we had to suppose
that only this mode is excited, thus restricting ourselves to a
certain polarization state of the incident light.
\begin{figure}[h]
\centering
\includegraphics[width=84mm,angle=0] {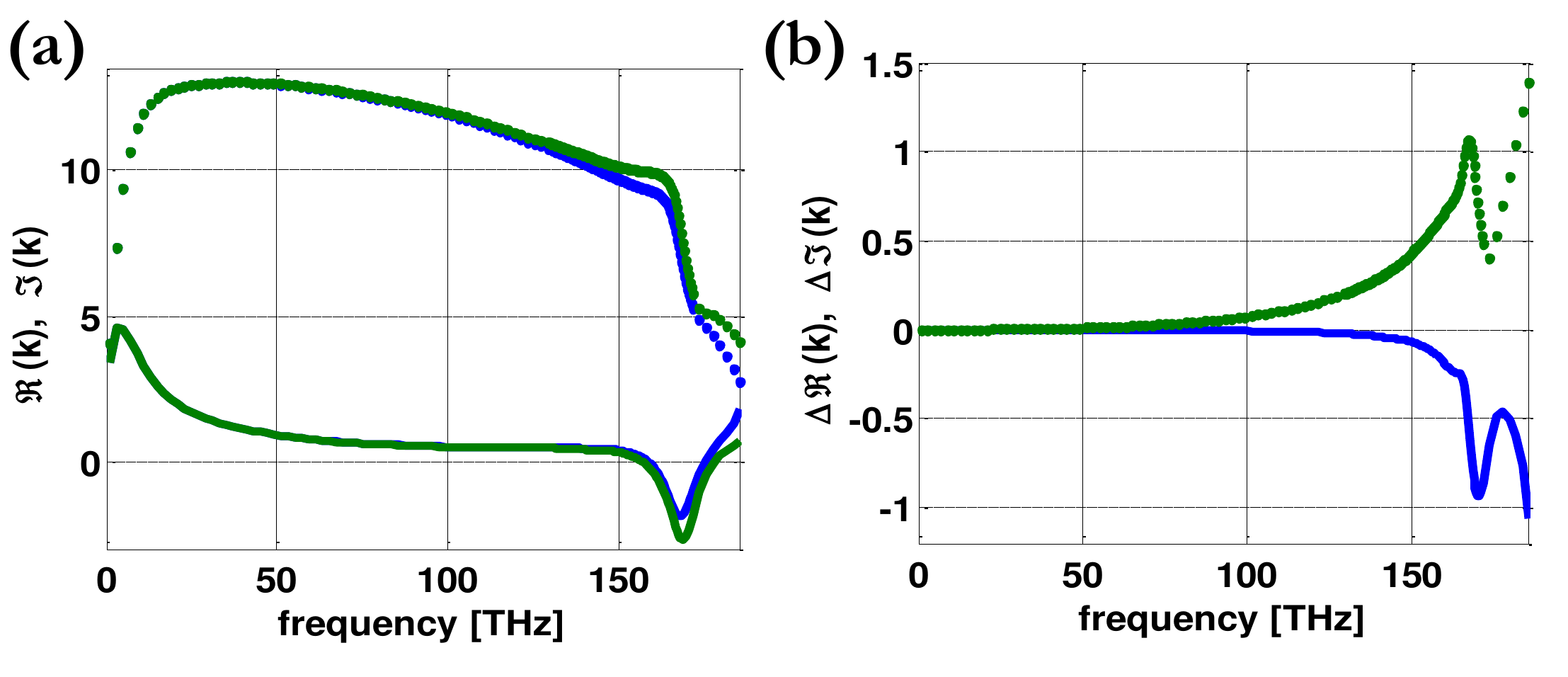}
\caption[Submanifold] {(color online) (a) Real part (solid line)
and imaginary part (dotted line) of the propagation constant $k
[\mu\mathrm{m}^{-1}]$ of the two lowest damped Bloch modes at
$k_\mathrm{t}=k_0$ that are degenerated at normal incidence. (b)
difference between the real parts (blue line) and the imaginary
parts (green line) of these two modes as function of the
frequency. In the quasi-static limit, i.e.
$\lambda\rightarrow\infty$ the difference is of course vanishing
and both modes are degenerated.} \label{FIG_KOS3}
\end{figure}
Nevertheless, in the general case of arbitrarily polarized
incident light one has to discuss both eigenstates of the
polarization separately, as both may exhibit different propagation
constants. Assuming an isotropic homogeneous medium the
eigensolutions are always degenerated, i.e. they have the same
propagation constant and it is possible to define the eigenmodes
as orthogonally and linearly polarized plane waves.
This is also the case for normally incident light on a $C_4$
symmetric structure as the isotropically designed SCiC
metamaterial. Here the light encounters the same physical
structure at normal incidence for any polarization, hence the
eigensolutions are degenerated. For oblique incidence the
situation changes dramatically because both linear polarized
eigenstates may encounter different structures. Both
eigensolutions may have different eigenvalues for the propagation
constant. One may compare the situation with an uniaxial crystal,
where the optical axis is aligned with the $z$-axis. For normally
incident light the structure can be considered as being isotropic,
whereas at oblique incidence the eigenvalues for the two
eigenpolarizations are different. Note that the eigenpolarizations
are always orthogonally, linearly polarized for planes of
incidence being identical with mirror planes of the metamaterial.

Considering the SCiC as an effectively isotropic homogeneous
medium, we expect that both the two fundamental Bloch modes are
degenerated and the effective refractive index is isotropic. So as
soon as the eigenvalues for both Bloch modes deviate from each
other, the medium cannot be described as being effectively
isotropic. This deviation can also serve as a measure for a
meaningful MM homogenization.

In Fig. \ref{FIG_KOS3} (a) we show the real and the imaginary part
of the propagation constant for the two Bloch modes with the
smallest losses at a certain angle of incidence and in Fig.
\ref{FIG_KOS3} (b) their difference as function of the frequency.
Clearly for low frequencies the propagation constants for both
modes are identical. As soon as the frequency increases the
propagation constants start to deviate from each other, hence the
SCiC cannot be considered as optically isotropic anymore. The
deviation is of course the stronger the larger the angle of
incidence is. At normal incidence they are identical as explained
in detail before. The conclusions to be drawn from these figures
are of course in line with those from the angular dependency of
the first Bloch mode only. The structure can be considered as
being isotropic only for small frequencies and a limited range of
angles of incidence where this angular range is the larger the
smaller the frequency is.

\section{Conclusion}
In this contribution we have investigated the optical response of
a highly symmetric SCiC  MM in the negative refraction regime. The
SCiC exhibits cubic symmetry, i.e. the highest possible symmetry
for periodic metamaterials, and is therefore considered a very
promising candidate for an isotropic negative refractive index
material. By investigating at first a finite number of functional
layers and the convergence of the corresponding effective
parameters to the values obtained from the dispersion relation of
the infinite structure we can conclude that the SCiC fulfills all
requirements of a homogenizable metamaterial. Already a single
functional layer can be described by its bulk properties as only a
single Bloch mode determines light propagation inside the
structure. Nevertheless our investigation of the iso-frequency
surface of the dispersion relation clearly shows that even this
metamaterial with the highest symmetry is far away from having an
optically isotropic response in the region of negative refraction.
Hence, the key result of this contribution is that one must not
conclude from high symmetry on an optically isotropic response.
Hence,  optical isotropy of a MM can be only deduced from a
careful inspection of the dispersion relation by taking into
account the material dispersion of metal too. This is an important
message to the designers of isotropic optical metamaterials.

\section*{Acknowledgements}
C.M., C.R., R.I and F.L.  acknowledge financial support by the
German Federal Ministry of Education and Research (Metamat) and by
the Thuringian State Government (MeMa). A.A., R.M. and A.L.
acknowledge partial supports from the Danish Research Council for
Technology and Production Sciences via the NIMbus project and from
COST Action MP0702.


\begin{thebibliography}{99}

\bibitem{Shelby2001} R. A. Shelby, D. R. Smith and S. Schultz, Science \textbf{292}, 77 (2001).

\bibitem{Smith2004} D. R. Smith, J. B. Pendry and M. C. K. Wiltshire, Science \textbf{305}, 788 (2004).

\bibitem{Dolling2006Science}G. Dolling, C. Enkrich, M. Wegener, C. M. Soukoulis and S. Linden ,Science \textbf{312}, 892 (2006).

\bibitem{Soukoulis2007Science} C. M. Soukoulis, S. Lindena and M. Wegener, Science \textbf{315}, 47 (2007).

\bibitem{Lezec2007} H. J. Lezec, J. A. Dionne and H. A. Atwater, Science \textbf{316}, 430 (2007).

\bibitem{Yao2008} J. Yao, Z. Liu, Y. Liu, Y. Wang, C. Sun, G. Bartal, A. M. Stacy, and X. Zhang, Science \textbf{321}, 930 (2008).

\bibitem{Pendry2000} J. B. Pendry, \prl \textbf{85}, 3966 (2000).

\bibitem{Balmaz2002} P. Gay-Balmaz and O. J. F. Martin, Appl. Phys. Lett. \textbf{81}, 939 (2002). 

\bibitem{Kussow2008} A. Kussow, A. Akyurtlu, and N. Angkawisittpan, Phys. Stat. Sol. B. \textbf{245}, 992-997 (2008). 

\bibitem{Vendik2006} I. Vendik, O. Vendik and M. Odit, Microwave and Optical Techn. Lett. \textbf{48}, 2553 (2006).  

\bibitem{Jelinek2006} L. Jelinek, J. Machac, and J. Zehentner, PIERS Online \textbf{2}, 624 (2006).  

\bibitem{Baena2006} J. D. Baena, L. Jelinek, R. Marqués and J. Zehentner, Appl. Phys. Lett. \textbf{88}, 134108 (2006) 

\bibitem{Simovski2003} C. R. Simovski, S. He, Phys. Lett. A \textbf{311}, 254 (2003). 

\bibitem{Verney2004} E. Verney, B. Sauviac, and C. R. Simovski, Phys. Lett. A \textbf{331}, 244 (2004). 

\bibitem{Baena2007} J. D. Baena, L. Jelinek, and R. Marqu\'{e}s, Phys. Rev. B. \textbf{76}, 245115 (2007). 

\bibitem{Enkrich2005}C. Enkrich, M. Wegener, S. Linden, S. Burger, L. Zschiedrich, F. Schmidt, J. F. Zhou, Th. Koschny and C. M. Soukoulis, \prl \textbf{95}, 203901 (2005). 

\bibitem{Gundogdu2006} T. F. Gundogdu, I. Tsiapa, A. Kostopoulos, G. Konstantinidis, N. Katsarakis, R. S. Penciu, M. Kafesaki and E. N. Economou, Appl. Phys. Lett. \textbf{89}, 084103 (2006). 

\bibitem{Menzel2009APL} C. Menzel, C. Helgert, J. \"Upping, C. Rockstuhl, E.-B. Kley, R. B. Wehrspohn, T. Pertsch and F. Lederer, Appl. Phys. Lett. \textbf{95}, 131104 (2009).

\bibitem{Menzel2009JOSAB} C. Menzel, R. Singh, C. Rockstuhl, W. Zhang, and F. Lederer, \josab \textbf{26}, B143 (2009).

\bibitem{Koschny2005}T. Koschny, L. Zhang, and C. M. Soukoulis, Phys. Rev. B. \textbf{71}, 121103(R) (2005).

\bibitem{Paul2009} T. Paul, C. Rockstuhl, C. Menzel and F. Lederer, \prb \textbf{79,} 115430 (2009).

\bibitem{Li2003PRE} Zhi-Yuan Li and Lan-Lan Lin, \pre \textbf{67}, 046607 (2003).

\bibitem{Andryieuski2009JOA} A. Andryieuski, C. Menzel, C. Rockstuhl, R. Malureanu and A. V. Lavrinenko, Journal of Optics A: Pure and Applied Optics \textbf{11}, 114010 (2009).

\bibitem{Andryieuski2009JEOS} A. Andryieuski, R. Malureanu, A. Lavrinenko, J. Europ. Opt. Soc. Rap. Public. \textbf{4}, 09003 (2009).

\bibitem{Dolling2006OL} G. Dolling, C. Enkrich, M. Wegener, C. M. Soukoulis and S. Linden, Opt. Lett. \textbf{31,} 1800 (2006). 

\bibitem{Pendry1996} J. B. Pendry, A. J. Holden, W. J. Stewart, I. Youngs, \prl \textbf{76}, 4773 (1996). 

\bibitem{Belov2003} P. A. Belov, R. Marqu\'{e}s, S. I. Maslovski, I. S. Nefedov, M. Silveirinha, C. R. Simovski and S. A. Tretyakov, \prb \textbf{67,} 113103 (2003) and Refs. therein. 

\bibitem{Hardy1981} W. N. Hardy, L. A. Whitehead, Rev. Sci. Instrum. \textbf{52}, 213 (1981). 

\bibitem{Linden2006Science} S. Linden, C. Enkrich, M. Wegener, J. Zhou, T. Koschny, C. M. Soukoulis, Science \textbf{306}, 1351 (2004). 

\bibitem{Penciu2008} R. Penciu, K. Aydin, M. Kafesaki, T. Koschny, E. Ozbay, E. Economou, and C. M. Soukoulis, Opt. Express 16, 18131 (2008).

\bibitem{SerdyukovBook} A. Serdyukov et al., \textit{Electromagnetics of Bi-Anisotropic Materials - Theory and Applications} (2001).

\bibitem{Rockstuhl2008PRB} C. Rockstuhl, C. Menzel, T. Paul, T. Pertsch and F. Lederer \prb \textbf{78,} 155102 (2008).

\bibitem{Smith2002} D. R. Smith, S. Schultz, P. Marko\v{s} and C. M. Soukoulis, \prb \textbf{65,} 195104 (2002).

\bibitem{Menzel2008PR} C. Menzel, C. Rockstuhl, T. Paul, F. Lederer and  T. Pertsch \prb \textbf{77}, 195328 (2008).

\bibitem{LiFMM} L. Li, \josaa {\bf 14,} 2758 (1997).

\bibitem{Rockstuhl2007PRB} C. Rockstuhl, T. Paul, F. Lederer, T. Pertsch, T. Zentgraf, T. P. Meyrath, and H. Giessen, \prb \textbf{77}, 035126 (2008).


\end{thebibliography}
\end{document}